\documentclass[onecolumn,showpacs,aps,prd,amsfonts,eqsecnum,
superscriptaddress,nofootinbib]{revtex4}
\usepackage{epsfig}
\usepackage{bm}

\begin{document}

\newcommand*{\nl}{\nonumber \\}
\newcommand*{\bea}{\begin{eqnarray}}
\newcommand*{\eea}{\end{eqnarray}}
\newcommand*{\bi}{\bibitem}
\newcommand*{\be}{\begin{equation}}
\newcommand*{\ee}{\end{equation}}
\newcommand*{\rms}{M_\rho^2(s)}
\newcommand*{\mrs}{m_\rho^2}
\newcommand*{\ra}{\rightarrow}
\newcommand*{\die}{e^+e^-}
\newcommand*{\eepppp}{e^+e^-\rightarrow\pi^+\pi^-\pi^+\pi^-}
\newcommand*{\rpppp}{\rho^0\rightarrow\pi^+\pi^-\pi^+\pi^-}
\newcommand*{\arp}{{a_1\rho\pi}}
\newcommand*{\eg}{e.g.}
\newcommand*{\rp}{{\rho^\prime}}
\newcommand*{\rpp}{{\rho^{\prime\prime}}}
\newcommand*{\amu}{{\mathbf A}^\mu}
\newcommand*{\anu}{{\mathbf A}^\nu}
\newcommand*{\amunu}{{\mathbf A}^{\mu\nu}}
\newcommand*{\vmu}{{\mathbf V}_\mu}
\newcommand*{\vmunu}{{\mathbf V}_{\mu\nu}}
\newcommand*{\fai}{\mathbf \phi}
\newcommand*{\rf}[1]{(\ref{#1})}
\newcommand*{\mas}{m_{a_1}^2}
\newcommand*{\lag}{{\mathcal L}}
\newcommand*{\p}{{\bm p}}
\newcommand*{\opava}{Institute of Physics, Silesian University in Opava,
Bezru\v{c}ovo n\'{a}m. 13, 746 01 Opava, Czech Republic}
\newcommand*{\praha}{Institute of Experimental and Applied Physics,
Czech Technical University, Horsk\'{a} 3/a, 120 00 Prague, Czech Republic}

\title{Electron--positron annihilation into four charged pions and the
$\bm{a_1\rho\pi}$ Lagrangian}
\thanks{This paper is dedicated to the late Julia Thompson,
who drew the attention of one of us (P.~L.) to the experimental program
of the Budker Institute of Nuclear Physics at Novosibirsk.}

\author{Peter Lichard}
\affiliation{\opava}
\affiliation{\praha}
\author{Josef Jur\'{a}\v{n}}
\affiliation{\opava}

\date{\today}% It is always \today, today,
             %  but any date may be explicitly specified

\begin{abstract}
The excitation curve of $e^+e^-$ annihilation into four charged pions
in the $\rho(770)$ region is calculated using three existing models with
$\rho$ mesons and pions in intermediate states supplemented by Feynman
diagrams with the $a_1(1260)\pi$ intermediate states. A two-term
phenomenological Lagrangian of the $a_1\rho\pi$ interaction is used.
The mixing angle is determined by fitting the
$e^+e^-\rightarrow\pi^+\pi^-\pi^+\pi^-$ cross section data of the
Novosibirsk CMD-2 collaboration and also its combination with the low-energy
part of the BaBar collaboration data. It is shown that the inclusion of the
$a_1\pi$ intermediate states succeeds in obtaining a good agreement with
the data on both cross section and the $\rho^0\rightarrow\pi^+\pi^-\pi^+\pi^-$  
decay width. When moving to energies above 1~GeV, the $\rho(1450)$ and
$\rho(1700)$ resonances are taken into account to get excellent agreement
with the BaBar data over the full energy range up to 4.5~GeV.
\end{abstract}

%14.40.Cs
\maketitle

\section{\label{sec:intro}Introduction}
Task of describing the excitation curve of the $\eepppp$ reaction
at low energies is closely related to the investigation of the energy 
dependent decay width of the four pion decay of $\rho(770)$. The validity 
of the factorization of the cross section into the $\rho(770)$ production
and decay parts is generally assumed on the basis of the vector meson
dominance (VMD) hypothesis \cite{vmd}. The assumption of factorization allows
experimentalists to determine a particular partial decay width at the nominal 
$\rho(770)$ mass on the basis of measurement the $\die$ annihilation cross 
section into the corresponding final state at the corresponding energy.
The current experimental value $\Gamma(\rpppp)=(2.8\pm1.4\pm0.5)$~keV was 
obtained in that way \cite{cmd2}.
In this work, we will also assume a one-to-one correspondence between
the cross section and the energy dependent decay width at the same energy, 
ignoring the complications which may appear if some conditions are not met 
\cite{app}. Needless to say that the evaluation of the decay width is 
less demanding technically and computationally than that of the cross 
section (five-dimensional quadrature instead of eight-dimensional one in 
the case of four-body final states).

The cross section of the $\die$ annihilation into the $2\pi^+2\pi^-$ and
$\pi^+\pi^-2\pi^0$ final states was considered by Decker, Heiliger, Jonsson,
and Finkemeier \cite{decker} in conjunction with the CVC-related decays 
of the $\tau$ lepton. The intermediate states of their model contained 
$\rho(770)$, $a_1(1260)$, and a scalar-isoscalar two-pion resonance.
In the two-charged-two-neutral case also the $\omega(782)$ was included.  
The $\arp$ vertex factors were adopted from the study of the three-pion
decay of the $\tau$ lepton by Isgur, Morningstar, and Reader \cite{isgur}. 
Czy\.{z} and K\"{u}hn \cite{czyz} constructed a Monte Carlo generator of 
the reaction
$\die\ra\gamma+4\pi$. The hadronic matrix elements were used in the form
suggested in \cite{decker}, corrected only for some minor deficiencies. To
check the soundness of their approach, Czy\.{z} and K\"{u}hn calculated 
also the excitation curves of the nonradiative reactions $\eepppp$ and 
$\die\ra\pi^+\pi^-2\pi^0$ in qualitative agreement with available data.
Ecker and Unterdorfer \cite{ecker3,ecker4} performed the first calculation
of the processes $\die\ra 4\pi$ and $\tau\ra\nu_\tau4\pi$ with the correct 
structure to $O(p^4)$ in the low energy expansion of the Standard
Model extrapolated to the resonance region. To get a good description of
the $\eepppp$ cross section up to 1~GeV, they had to include as additional
contribution the $a_1$ exchange. They circumvented the $a_1\rho\pi$ Lagrangian
ambiguity by choosing a special relation among the individual coupling
constants.

The four-pion decays of the $\rho(770)$ are generally considered
a convenient test ground of the low-energy effective theories of the
interactions of $\rho$ mesons and pions. In the past, several papers
appeared that calculated the corresponding partial decay widths
\cite{bramon,eidelman,plant,achasov1,achasov2}. Moreover, Achasov
and Kozhevnikov \cite{achasov1,achasov2} argued that the four-pion decay
widths of $\rho(770)$ are not experimentally well defined because they
require the
averaging over a mass interval in which they rise rapidly. They therefore
calculated, in addition to the decay widths $\Gamma(\rpppp)$,
$\Gamma(\rho^0\ra\pi^+\pi^-\pi^0\pi^0)$, $\Gamma(\rho^+\ra\pi^+3\pi^0)$, and
$\Gamma(\rho^+\ra2\pi^+\pi^-\pi^0)$, the $\eepppp$ reaction cross
section as a function of incident energy (excitation curve) and compared
it to the CMD-2 data \cite{cmd2} from Novosibirsk.

With respect to the role of the axial-vector meson $a_1(1260)$ in the
four-pion decays of $\rho^0$, the situation is somewhat controversial. 
On one side, the intermediate states containing the $a_1(1260)$ were 
either ignored \cite{bramon,eidelman,plant,achasov1,achasov2} or shown 
\cite{plant} to have little influence on the four-pion decay widths
of $\rho(770)$\footnote{Ecker and Unterdorfer \cite{ecker3,ecker4} also
included the $a_1$ contribution when calculated the $\rho^0\ra 4\pi$
branching fractions, but it is impossible for us to assess its role because
they did not show results without it.}.  
On the other side, the analysis of the differential
distributions of charged pions coming from the $\die$ annihilation
in the energy range 1.05--1.38 GeV demonstrated the dominance
of the $a_1(1260)\pi$ intermediate states \cite{akhmetshin1999}.
Given the large width of $a_1(1260)$ ($\Gamma_{a_1}$=250 to 600 MeV
\cite{pdg2006}) it would be surprising if the
role of the $a_1$-meson diminished so fast outside the above range, which is
not too far from the $\rho(770)$ mass. Moreover, the $a_1(1260)$ meson
was shown to be important in the four-pion decays of the $\tau$-lepton
\cite{decker,bondar1999,edwards2000}, which are in a sense isospin
counterparts of the four-pion final states in the $\die$ annihilation.
Recently, the paper of Achasov and Kozhevnikov has appeared \cite{achasov3}
that included the intermediate states with the $a_1$ meson using the
generalized hidden local symmetry model \cite{bando1988}. This increased the
$\rho^0\rightarrow 2\pi^+2\pi^-$ decay width from 0.94~keV to 1.59~keV
assuming the nominal mass of the $a_1$ resonance, see Table~I in
\cite{achasov3}. Unfortunately, the authors of \cite{achasov3} do not provide 
the comparison with the $e^+e^-\rightarrow 2\pi^+2\pi^-$ cross section,
which has a greater discriminatory value than the decay width
alone \cite{achasov1,achasov2}.

The present work was triggered by our interest in the electromagnetic 
probes in relativistic heavy-ion collisions.
The production of prompt dileptons and photons is for a long time considered
a powerful tool for investigating the properties of dense systems created
in the hadronic and nuclear collisions \cite{feinberg, shuryak}. The interest
of the heavy ion community in dilepton production has recently been boosted
by very precise dimuon data by the CERN/NA60 collaboration \cite{na60}. The
prompt dileptons and photons can, in principle, originate from two sources:
(i) quark-gluon plasma, (ii) hadron gas. The theoretical calculations of the
dilepton and photon yield from the latter are hampered by not uniquely known
Lagrangian of the $\arp$ interaction \cite{song,gaogale}. We suggest one way
how to relieve this problem and make the predictions of the dilepton and
photon production from hadron gas more reliable. The results of the present
work have already been utilized \cite{joerg} in the evaluation of the dimuon 
production rate in In-In collisions at 158 A GeV.

A way of narrowing the uncertainty interval in dilepton production from 
the hadron
gas was advocated a long time ago by Li and Gale \cite{ligale}. They checked
the feasibility of the dilepton rate calculation in, \eg, $\omega\pi^0$
collisions, by comparing the inverse process ($\die$ annihilation into the
$\omega\pi^0$ final state) with the available experimental data. Li and Gale
also considered the $\die$ annihilation into four pions in the narrow
$a_1$ width approximation, i.e. as process $\die\ra a_1\pi$. In the
present work we handle it as a genuine 
four-final-pion process, considering not only the $a_1\pi$ intermediate
states but also other intermediate states following from various
models based on the chiral perturbation theory.

\begin{figure}
\setlength \epsfxsize{8.6cm}
\epsffile{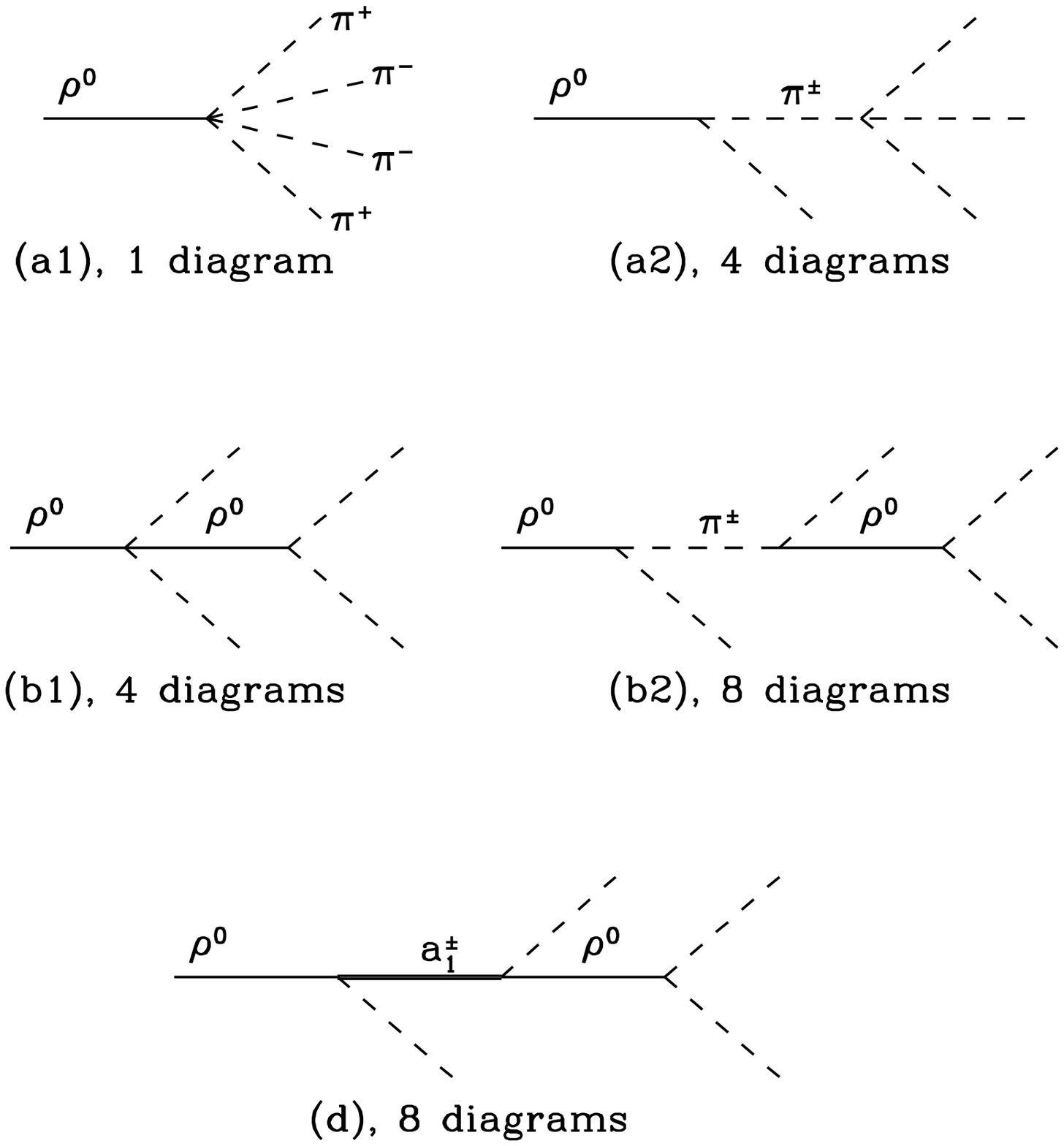}
\caption{\label{fig:diagrams}Selected Feynman diagrams describing
decay $\rpppp$}
\end{figure}

In this paper we investigate the role of the $a_1(1260)$ resonance
in the $\die$ annihilation into four charged pions and in the 
four-charged-pion decays of $\rho(770)$ in more detail. We supplement
three existing models, which consider only $\rho$ and $\pi$ in intermediate
states (diagrams (a) and (b) in Fig.~\ref{fig:diagrams}), with the $a_1$
contribution (diagrams (d) in Fig.~\ref{fig:diagrams}). Those three models
are (1) the model of Eidelman, Silagadze, and Kuraev \cite{eidelman},
(2) one of the models considered by Plant and Birse \cite{plant}, and
(3) the model of Achasov and Kozhevnikov \cite{achasov1,achasov2}.
We will consider only the final state with all charged pions, for which
the experimental data are best. The difference between the approach of 
Achasov and Kozhevnikov \cite{achasov3} and ours lies mainly in the
$\arp$ Lagrangian. In \cite{achasov3}, the generalized hidden
local symmetry Lagrangian was chosen, whereas we choose a more
phenomenological two-term Lagrangian, which is often used in a different
branch of the particle physics. Its individual terms and their specific 
combinations appeared in many papers computing the dilepton 
and photon production rate from a thermalized meson gas. In this paper 
we consider the mixing angle between the two terms as a free parameter and 
will determine its value by fitting the excitation curve of the 
$\eepppp$ reaction.

In order to evaluate the amplitude induced by eight diagrams
Fig.~\ref{fig:diagrams}(d) we have to
choose a Lagrangian of the $\arp$ interaction. For this choice, there are
basically two approaches in the literature. One explores well defined
theoretical concepts to build dynamical models, the free parameters of
which are then fixed by comparison with observed masses and decay widths.
See, \eg, 
Refs.~\cite{wess,gomm84,holstein,meissner,kaiser,song,korudaz,li,smejkal97}.

In  other, more phenomenological, approaches the authors simply chose
for the $\arp$ Lagrangian various expressions built from the field operators 
and compatible with the fundamental conservation laws. Such Lagrangians, 
after fixing their coupling constants, were then used to calculate various 
observable quantities, see, \eg, \cite{pham,kuhnsanta,janssen,haglin}.
In some approaches \cite{isgur,xiong}, directly the vertex factors 
were written without showing the corresponding Lagrangian. From the fact 
that different authors pick different Lagrangians one might get the 
impression that the choice of Lagrangian 
is not very important and that various Lagrangians lead to identical, or at
least similar, results. This is not true, and the observable
quantities may be very sensitive to the choice of the $\arp$ Lagrangian, as
was demonstrated, \eg, by Song \cite{song}. The discussion of the
$a_1$ phenomenology with emphasis on the photon and dilepton production
from a hot meson gas can be found in Ref. \cite{gaogale}.

With so many $\arp$ effective Lagrangians, it is
interesting to learn which Lagrangian is preferred experimentally. This
would require constructing a general Lagrangian with a set of free
parameters and fixing them by comparing all possible observables with
the existing data. Of course, such a program is very ambitious. In this
paper we are going to do something much simpler. Below, we choose a
two-component Lagrangian and determine its two free parameters by requiring
that the decay width of $a_1(1260)$ be reproduced and the best possible fit
obtained for the excitation curve of the $\eepppp$ reaction. Even
this restricted program cannot be accomplished completely. Firstly, the
width of $a_1(1260)$ is not known reliably. We will consider three values
from the range given in the Particle Data Group tables \cite{pdg2006},
namely 250, 400, and 600 MeV. Secondly, the result of the fit will depend
also on the basic $\rho$ and $\pi$ intermediate state model to which we add
the $a_1\pi$ contribution. Nevertheless, we will show that the inclusion of
the $a_1\pi$ intermediate states is necessary for obtaining good agreement
with the $\eepppp$ excitation curve.

Following the suggestion of T.~Barnes \cite{barnes} we also include the 
ratio of the $D$-wave and $S$-wave amplitudes of the $a_1\ra\rho\pi$ decay 
as a fitted quantity. The importance of the $D/S$ ratio for
selecting among the $\arp$ Lagrangians was stressed in a different context
in \cite{gaogale}. We use the experimental value $D/S=-0.14\pm0.11$
\cite{DSratio}, which was obtained by genuine partial wave analysis of the
reaction $\pi^- p\ra\pi^+\pi^-\pi^- p$. We consider the other values which 
exist in literature strongly model biased. They were obtained by fitting
the three-pion mass spectrum in the decay $\tau\ra\nu_\tau 3\pi$ using
the model of Isgur, Morningstar, and Reader \cite{isgur} and then
calculating the $D/S$ ratio from the optimal parameters of the model.

\section{\label{sec:models}Original models, modifications, and additions}
As we already stated, we will complement three existing models of the
four-pion decays, which consider only intermediate states with $\rho$ mesons
and pions, with the intermediate states containing the axial-vector
resonance $a_1(1260)$. We must admit that our choice of models is rather
arbitrary. Moreover, we took the models as they appeared in the
literature and did not check their compatibility with
chiral symmetry and with the constraints on coupling of resonances to
the pions \cite{ecker1,ecker2}.
 
Those three models are characterized below.

\subsection{Model of Eidelman, Silagadze, and Kuraev (ESK)}
This model \cite{eidelman} is based on the effective chiral Lagrangian
by Brihaye, Pak, and Rossi \cite{brihaye}, which was investigated also
in \cite{kuraev}. It follows from that Lagrangian that all (a) and (b)
diagrams depicted on Fig.~\ref{fig:diagrams} contribute to the $\rpppp$
decay rate. Their amplitudes (in the notation slightly different from ours)
are shown in the paper. Our
usage of this model will differ from the original paper in three respects:
(1) We add $a_1$ diagrams Fig.~\ref{fig:diagrams}(d). (2) We use a
different value of the parameter $\alpha_k$, defined in \cite{eidelman}.
Instead of 0.55 we set $\alpha_k=0.5$,
which follows from the KSRF relation \cite{ksrf}, to be in conformity with
other two models. (3) We replace the scalar part of the $\rho$-meson 
propagator with fixed mass and fixed width by the prescription
\be
\label{rhoprop} P_\rho(s)=-\frac{i}{s-\rms+i m_\rho\Gamma_\rho(s)},
\ee
which uses the running mass squared $\rms$ and the energy dependent total
width $\Gamma_\rho(s)$ from Ref. \cite{ratio}.

The last point deserves more comments. The denominator of our propagator
\rf{rhoprop} is an analytic function in the $s$-plane with a cut running
from $4m_\pi^2$ to infinity, as required by general principles. This
property differs \rf{rhoprop} from most of the formulas used in the
literature. The real function $\rms$ is calculated from $\Gamma_\rho(s)$
using a once subtracted dispersion relation, which guarantees that the
condition $M_\rho^2(\mrs)=\mrs$ is satisfied. Further condition
\be
\label{rmsderiv}
\left.\frac{d \rms}{d s}\right|_{s=\mrs}=0
\ee
is not fulfilled automatically and
serves as a test that all important contributions to the total $\rho$-meson
width $\Gamma_\rho(s)$ have properly been taken into account. See \cite{ratio}
for details.
If we replace the $m_\rho$ accompanying $\Gamma_\rho(s)$ in Eq.~\rf{rhoprop}
by $\sqrt{s}$, as it is done in some existing formulas, the condition
\rf{rmsderiv} cannot be satisfied for any reasonable choice of
$\Gamma_\rho(s)$. 

The running mass approach \cite{ratio} takes into account, in addition
to the basic two-pion decay channel, several channels 
($\omega\pi^0$,  $K^+K^-$, $K^0\bar K^0$, and $\eta\pi^+\pi^-$) which open 
as the $\rho$ resonance goes above its nominal mass. It also considers 
structure effects described by the strong form factors. In these two respects 
it differs from other approaches that appeared in the literature 
\cite{GS,VW,melikhov04}. Gounaris and Sakurai \cite{GS} considered only 
the two-pion contribution to the total width of the $\rho^0$ resonance 
and ignored structure effects. Vaughn and Wali \cite{VW} took into account 
the strong form factor, but again ignored higher decay channels. Melikhov, 
Nachtmann, Nikonov, and Paulus \cite{melikhov04} included the $K^+K^-$ 
and $K^0\bar K^0$ channels, but did not consider the strong form factors.

We will use the $\rho$ propagator \rf{rhoprop} not only in conjunction with
the ESK model, but also with the other ones. This is the main reason why our
results calculated within the original models (i.e., without the $a_1\pi$
intermediate states) and presented below differ slightly from the results
quoted in the original papers.

\subsection{One of the models of Plant and Birse (PB/HG)}
Plant and Birse \cite{plant} investigated several models of the four-pion 
decays of
$\rho^0(770)$. One of them (labeled HG) is a corrected version
of the model by Bramon, Grau, and Pancheri \cite{bramon}, which was
based on the hidden gauge theory of Bando \textit{et al.}
\cite{bando1985}. The 2$\pi$2$\rho$ contact terms, see diagram (b1)
in Fig.~\ref{fig:diagrams},
are missing in this approach. The amplitude of the (a1) diagram is different
from that in work by Eidelman, Silagadze, and Kuraev \cite{eidelman} by
a factor $(-1/2)$. The amplitudes of (a2) and (b2) diagrams are equal to
their ESK counterparts. 

\subsection{Model of Achasov and Kozhevnikov (AK)}
Achasov and Kozhevnikov \cite{achasov1,achasov2} studied the four-pion
decays of $\rho(770)$, five-pion decays of $\omega(782)$, and the processes
related to them. Namely, the $\die$ annihilation into the four- and five-pion
final states and the four-pion decays of the $\tau$ lepton. They
used the Weinberg Lagrangian \cite{weinberg} obtained upon the nonlinear
realization of chiral symmetry. From their rather extensive work we adopt
their prescriptions for the amplitudes (a1), (a2), and (b2) of the
$\rpppp$ decay. The contact amplitude (b1) is again vanishing.
Achasov and Kozhevnikov used a fixed-mass, variable-width formula for the
$\rho$-meson propagator.

\subsection{$\bm{\arp}$ Lagrangian and the amplitude of the diagrams
containing the $a_1$ meson}
We choose the following interaction among the $a_1$, $\rho$, and $\pi$
fields
\be
\label{genlag} \lag=\frac{g_{\arp}}{\sqrt{2}}
\left(\lag_1\cos\theta+\lag_2\sin\theta\right),
\ee
where $g_{\arp}$ and $\theta$ are yet undetermined parameters,
\bea
\label{lag1}
\lag_1&=& \amu\cdot\left(\vmunu\times\partial^\nu{\fai}\right),\\
\label{lag2}
\lag_2&=& \vmunu\cdot\left(\partial^\mu\anu\times{\fai}\right),
\eea
and $\vmunu=\partial_\mu{\mathbf V}_\nu-\partial_\nu\vmu$. The isovector
composed of the $\rho$-meson field operators is denoted by $\vmu$, similar
objects for $\pi$ and $a_1$ are $\fai$ and $\amu$, respectively. We
write
\bea
\phi_1&=&\frac{1}{\sqrt 2}\left(\phi_c+\phi_c^\dagger\right),\nl
\phi_2&=&\frac{i}{\sqrt 2}\left(\phi_c-\phi_c^\dagger\right),\nl
\phi_3&=&\phi_n, \nonumber
\eea
and assume that $\phi_c$ contains the annihilation operators of the
positive pion and creation operators of the negative pion. $\phi_n$ is
the operator of neutral pion field.

A specific combination of terms \rf{lag1} and \rf{lag2} appeared in the
pioneering work by Wess and Zumino \cite{wess}. Term \rf{lag1} alone was
used by Xiong, Shuryak, and Brown \cite{xiong} in their study of the photon
production from meson gas. Janssen, Holinde, and Speth \cite{janssen}
picked the term \rf{lag2} when they evaluated the amplitude of the $\pi\rho$
scattering. Another combination of \rf{lag1} and \rf{lag2} appeared in
the calculation of dilepton production from meson gas by Song, Ko, and Gale
\cite{sokoga}.

Lagrangian \rf{genlag} leads to the following factor for
the vertex in which an incoming $a_1^+$ (index $\alpha$), an outgoing
$\rho^0$ (index $\mu$), and an outgoing $\pi^+$ meet
\bea
V^{\alpha\mu}\left(p_{a_1},p_\rho,p_\pi\right)&=&\frac{g_{a_1\rho\pi}}
{\sqrt{2}}\left\{\cos\theta
\left[p_\rho^\alpha p_\pi^\mu-(p_\pi p_\rho)g^{\alpha\mu}\right]\right.\nl
&-&\left.\sin\theta
\left[p_\rho^\alpha p_{a_1}^\mu-(p_{a_1}p_\rho)g^{\alpha\mu}\right]\right\}.
\label{vertex}
\eea
The $a_1^-\rho^0\pi^-$ vertex acquires an extra minus sign.
The evaluation of the decay rate of $a_1^+\ra\rho^0\pi^+$ using vertex
\rf{vertex} is straightforward.
\bea
\Gamma_{a_1^+\ra\rho^0\pi^+}&=&\frac{g^2_{\arp}}
{192\pi m_{a_1}^3}
\lambda^{1/2}(m_{a_1}^2,m_\rho^2,m_{\pi^+}^2)\nl
&\times&R(m_{a_1}^2,m_\rho^2,m_{\pi^+}^2)\ ,
\label{gamrhopiplus}
\eea
where
\[
\lambda(x,y,z)=x^2+y^2+z^2-2xy-2xz-2yz
\]
and
\begin{eqnarray*}
R(x,y,z)&=&\left[(x-y-z)^2+\frac{y}{2x}(x-y+z)^2\right]\cos^2\theta\\
&-&2\left[(x-z)^2+y(x+z-2y)\right]\cos\theta\sin\theta\\
&+&\left[(x+y-z)^2+2xy\right]\sin^2\theta\ .
\end{eqnarray*}
If we assume the charge independent $\arp$ coupling constant and masses
of $\rho$ and $a_1$, the width of the decay $a_1^+\ra\rho^+\pi^0$ is
obtained from \rf{gamrhopiplus} by changing just the pion mass. 
Formula 
\be
\Gamma_{a_1^+}=\Gamma_{a_1^+\ra\rho^0\pi^+}+\Gamma_{a_1^+\ra\rho^+\pi^0}
\label{gamrhopi}
\ee
 enables us to find the coupling constant
$g_{\arp}$ for given $\Gamma_{a_1^+}$ and $\sin\theta$.
Because each of the diagrams Fig.~\ref{fig:diagrams}(d) contains two $\arp$
vertices, the overall sign of the Lagrangian \rf{genlag} is not important
and we can assume a non-negative $\cos\theta$.

A question arises whether the narrow $\rho$-width approximation used above 
is accurate enough for the purpose of determination the $g_{\arp}$ coupling 
constant. Achasov and Kozhevnikov \cite{achasov3} showed that the $a_1$ 
decay width calculated as $\Gamma(a_1\ra\rho\pi)$  came out larger 
than that calculated as $\Gamma(a_1\ra 3\pi)$ using the same coupling 
constant $g_{\arp}$. We have examined this issue, too, using our two-component 
Lagrangian \rf{genlag} and got that the
$\Gamma(a_1\ra 3\pi)/\Gamma(a_1\ra\rho\pi)$ ratio is smaller than unity
(like in \cite{achasov3}) for $\sin\theta\lesssim 0.2$ and 
$\sin\theta\gtrsim 0.65$. In the remaining interval, which contains all the 
values of $\sin\theta$ that will be met in our calculations, this ratio
is greater than one. Moreover, we have found that if one takes into account 
the strong form factors in the $\arp$ and $\rho\pi\pi$ vertices, the results 
of both approaches become almost identical. This can be explained as
follows. According to Kokoski and Isgur \cite{kokoski} formula, which will 
be shown later, the strong form factor in a particular decay vertex is 
a decreasing function of the three-momentum of an outgoing particle in the 
rest frame of the parent particle. In the $a_1\ra3\pi$ decay the intermediate 
mass of $\rho$ is mostly smaller than its nominal mass, what means higher 
momenta of particles emerging from the $\arp$ vertex. In addition, the 
$a_1\ra3\pi$ width is further reduced by the form factor in the
$\rho\pi\pi$ vertex.

Let us now turn to the amplitude of the $a_1$ diagrams
Fig.~\ref{fig:diagrams}(d) for the decay
$\rho^0(p)\ra\pi^-(p_1)\pi^+(p_2)\pi^+(p_3)\pi^-(p_4)$. We first
introduce the notation
\bea
q_i &=& p-p_i, \nl
r_{ij}&=&p_i+p_j, \nl
s_{ij}&=&r_{ij}^2,
\label{sij}
\eea
and then write the amplitude in the form
\[
{\mathcal M}_d^{(\lambda)}=\epsilon_\lambda^\mu J_{d,\mu},
\]
where $\epsilon_\lambda^\mu$ is the polarization vector of the
decaying $\rho^0$ and
\begin{eqnarray*}
J_{d,\mu}&=&-(1-P_{12}P_{34})(1+P_{14})(1+P_{23})V_{\alpha\mu}(-q_4,-p,p_4)\\
&\times&P_{a_1}^{\alpha\beta}(q_4)V_{\beta\nu}(q_4,r_{12},p_3)
g_\rho(p_2-p_1)^\nu P_\rho(s_{12}).
\end{eqnarray*}
Here, $P_{ij}$ denotes the operator that interchange four-momenta $p_i$
and $p_j$. The axial-vector meson propagator
\be
\label{a1propagator}
P^{\alpha\beta}_{a_1}(q)=i\frac{-g^{\alpha\beta}+\frac{1}{m_{a_1}^2}
q^\alpha q^\beta}
{q^2-m_{a_1}^2+i m_{a_1}\Gamma_{a_1}}
\ee
is chosen in a simple fixed-mass, fixed-width form. Here, we are going
to consider the four-pion system with invariant energies less than 1~GeV.
The invariant mass of the three-pion system, which is equal to $\sqrt{q^2}$,
is thus limited by 0.86~GeV. Achasov and Kozhevnikov \cite{achasov3} showed
that the $a_1$ decay width is negligible in that energy range. Referring
to their finding we set $\Gamma_{a_1}=0$ in Eq.~\rf{a1propagator}.
The scalar part of the $\rho$ propagator is again used in the form
\rf{rhoprop}.

\subsection{Technicalities}

The complete amplitude of the $\rpppp$ decay is
\be
\label{ampl}
{\mathcal M}^{(\lambda)}=\epsilon_\lambda^\mu J_\mu,
\ee
where $\epsilon_\lambda^\mu$ is the polarization vector of the
decaying $\rho^0$ and
\[
J_\mu=J_{a,\mu}+J_{b,\mu}+J_{d,\mu}.
\]
Four-vectors $J_{a,\mu}$ and $J_{b,\mu}$ describe the contributions from (a)
and (b) diagrams in a
particular model and $J_{d,\mu}$ is the contribution from (d) diagrams.
The sum over the $\rho$-meson polarizations of the amplitude \rf{ampl} squared is given by
\be
\label{amplsq}
\sum_\lambda\left|{\mathcal M}^{(\lambda)}\right|^2
=\left(-g^{\mu\nu}+\frac{p^\mu p^\nu}{m_\rho^2}\right)
J_\mu J_\nu^*.
\ee

This formula is more complicated than that used in \cite{eidelman}, because
the four-vectors $J_{a,\mu}$ and $J_{b,\mu}$ of PB/HG and AK models do
not satisfy the transversality condition $J_\mu p^\mu=0$. We used the
algebraic manipulation program {\textsc REDUCE} \cite{reduce} to express
the sum \rf{amplsq} in terms of six invariants $s_{ij}$, $i<j$, $j=2,3,4$
defined in \rf{sij}. Of course, only five of them are independent and
we used the identity $\sum_{i<j}s_{ij}=m_\rho^2+8m_\pi^2$ for the checks
in the process of evaluation of the decay width. When calculating the
excitation curve, $m_\rho^2$ is replaced by $s$, the square of the incident
energy.

When calculating the decay width of an unpolarized parent particle,
we may take advantage of the spherical symmetry of the problem and choose
the following kinematic configuration: (1) The parent particle $a$ is at
rest. (2) The summed momentum $\p_{12}$ of particles 1 and 2 points in the
direction of the $z$ axis. (3) The individual momenta $\p_1$ and $\p_2$ lie
in the $xz$ plane. Then the following formula, written in a general case
with arbitrary masses and spins, is valid
\bea
\label{gamma4}
 \Gamma &=& \frac{N}{16(2\pi)^6m_a^2}\int_{m_1+m_2}^{m_a-m_3-m_4} d m_{12} \
 p_1^*\int_{m_3+m_4}^{m_a-m_{12}} d m_{34}\nl &\times& p_{12} p_3^*
 \int_{-1}^1 d\cos\theta_1^*\int_{-1}^1 d\cos\theta_3^* \int_0^{2\pi} d \varphi_3
 \overline{\left|{\mathcal M}\right|^2}.
\eea
The last quantity is the amplitude squared, averaged over the initial
spin states, and summed over the final spin states. The factor $N$ takes into
account the identity of the final particles and equals 1/4 in our case.
The asterisk denotes the momentum in the corresponding rest frame (1-2 or 3-4),
$p=|\mathbf{p}|$, and $m_{ij}=\sqrt{s_{ij}}=\sqrt{(p_i+p_j)^2}$.

For evaluation of the integrals in \rf{gamma4} we used a sequence of the
five one-dimensional Gauss-Legendre quadratures of the sixteenth order. 
We prefer this method to Monte-Carlo integration because we use the result 
of the integration in a minimization procedure and therefore we require that
the same value of the optimized variable (Lagrangian mixing angle) yield
always the same value of the minimized function, which would not be
satisfied with the Monte-Carlo integration. Nevertheless, we checked our
computer code by evaluating the decay width for a particular value of the
mixing angle using a completely independent code based on the Monte-Carlo
method.

 To convert the calculated decay width into the cross section, we start
with the formula
\[
\sigma_{4\pi}(s)=\frac{\sigma_{\pi^+\pi^-}(s)}{\Gamma_{\pi^+\pi^-}(s)}
\Gamma_{4\pi}(s).
\]
Using
\[
\sigma_{\pi^+\pi^-}(s)=\frac{\pi\alpha^2}{3s}
\left(1-\frac{4m_\pi^2}{s}\right)^{3/2}\left|F_\pi(s)\right|^2,
\]
where $F_\pi(s)$ is the contribution of the $\rho$ resonance to the
pion form factor, and
\[
\Gamma_{\pi^+\pi^-}(s)=\frac{g_\rho^2W}{48\pi}
\left(1-\frac{4m_\pi^2}{s}\right)^{3/2}
\]
with $W=\sqrt{s}$, we arrive at
\be
\sigma_{4\pi}(s)=
\left(\frac{4\pi\alpha}{g_\rho}\right)^2\frac{1}{W^3}
\left|F_\pi(s)\right|^2\Gamma_{4\pi}(s).
\label{convert}
\ee
We further use the VMD expression for the dielectron
decay width of $\rho^0$
\be
\Gamma_{\die} = \frac{4\pi m_\rho}{3}\left(\frac{\alpha}{g_\rho}\right)^2
\label{rhoee}
\ee
and get
\be
\label{aksigma}
\sigma_{4\pi}(s)=
\frac{12\pi\Gamma_{\die}}{m_\rho W^3}
\left|F_\pi(s)\right|^2 \Gamma_{4\pi}(s).
\ee
If we set, following Achasov and Kozhevnikov \cite{achasov2},
\be
\label{akpiff}
F_\pi(s)=\frac{m_\rho^2}{D_\rho(s)},
\ee
where the inverse $\rho$-meson propagator $D_\rho(s)$ is defined in
Eq.~(2.2) of \cite{achasov2}, we reproduce their Eq.~(3.1).

In our opinion, Eq.~\rf{aksigma} overestimates the
cross section if the experimental value of $\Gamma_{\die}$ is used.
The reason is that
the dielectron decay width calculated from \rf{rhoee} is smaller than
the experimental value. We will therefore stick with formula \rf{convert}.

We utilize the scalar part of the $\rho$-meson propagator
\rf{rhoprop} to write our Ansatz for the $\rho$-meson contribution to the
pion form factor
\be
\label{piff}
F_\pi(s)=\frac{M_\rho^2(0)}{M_\rho^2(s)-s-im_\rho\Gamma_\rho(s)}.
\ee
As shown in \cite{ratio}, this formula gives the correct value of the
mean square radius of the pion. The form factor \rf{akpiff}
fails in this test.

For the $\rho$ coupling constant we use the same value as in
\cite{plant,achasov1,achasov2}, namely $g_\rho=5.89$. This value is 
compatible with what follows from the KSRF relation \cite{ksrf} 
($5.900\pm0.011$). Both values are little lower than $g_\rho=6.002\pm0.015$ 
calculated from the $\rho$-meson width.

\section{Low-energy results ($\bm{W<1\ \mathrm{GeV}}$)}
\label{lowenergy}
We deal with the excitation curves of the reaction $\eepppp$ calculated
in three different models (ESK, PB/HG, AK) supplemented with the $a_1$
diagrams Fig.~\ref{fig:diagrams}(d). We first fit them to the CMD-2 data
\cite{cmd2} by varying
the sine of the mixing angle $\theta$, defined in \rf{genlag}, for the
three fixed values of the width of the $a_1(1260)$ meson. We did not
consider the first two points in the CMD-2 data, because they give only
upper bounds of the cross section. The ratios of the usually defined 
$\chi^2$ to the
number of degrees of freedom (NDF), which characterize the quality
of the fit, are shown in Table~\ref{tab:chiscmd2}. The last row in
Table~\ref{tab:chiscmd2} shows the values of $\chi^2$/NDF that indicate 
how well (or badly) the original
models without $a_1$ agree with the data. No free parameter is involved
in the latter case. Table~\ref{tab:chiscmd2} shows that
the ratio $\chi^2/$NDF is always greater than one. In what
follows, we shall therefore multiply the statistical errors of the 
quantities obtained in the process of minimization by the square root of
that ratio. 
\begin{table}
\caption{\label{tab:chiscmd2}$\chi^2/$NDF of the fits to the CMD-2 cross
section data (11 data points)}
\begin{ruledtabular}
\begin{tabular}{ccccc}
$ \Gamma_{a_1}$ & ESK  & PB/HG & AK   & only $a_1$ \\
 (MeV) & \cite{eidelman} & \cite{plant} & \cite{achasov1,achasov2} & \\
\hline
250 &   1.60   &   1.34   &   1.28   &   1.68   \\
400 &   1.53   &   1.37   &   1.30   &   1.82   \\
600 &   1.61   &   1.41   &   1.31   &   1.94   \\
\hline
 Only $\rho$, $\pi$& 17.6  & 15.0 & 14.8 &  / \\
\end{tabular}
\end{ruledtabular}
\end{table}

 The inspection of Table~\ref{tab:chiscmd2} shows that the inclusion of
the $a_1$ contribution greatly improves the agreement with the data.  The
interference between the original diagrams and the new ones is important,
the results of the combined model are better than those of the $a_1$ diagrams
alone. The best results (lowest $\chi^2$) are obtained with the AK model 
supplemented with the $a_1\pi$ intermediate states.

To investigate the sensitivity of our results to the input data,
we combine the CMD-2 data \cite{cmd2} and the low-energy ($s<1$~GeV$^2$) 
part of the BaBar data \cite{babar} (BaBar-LE in what follows) into a new 
set and repeat the calculations.
The results are shown in Table~\ref{tab:chiscomb}. Their comparison
with the results obtained from the CMD-2 data alone shows two important
differences: (i) The agreement of all models with data, characterized 
by $\chi^2$/NDF, is now better. It indicates that the CMD-2 and BaBar-LE 
data are compatible, so the increased number of data points does not bring
proportional increase of $\chi^2$. (ii) Whereas merging of the $a_1$
diagrams with the PB/HG or AK model improves the fit, adding the ESK model
diagrams to the pure $a_1$ contribution leads to the opposite effect.

\begin{table}
\caption{\label{tab:chiscomb}$\chi^2$/NDF of the fits to
the combined CMD-2 \& BaBar-LE cross section data (27 data points)}
\begin{ruledtabular}
\begin{tabular}{ccccc}
$ \Gamma_{a_1}$ & ESK   & PB/HG  & AK   & only $a_1$ \\
 (MeV)        &\cite{eidelman} &\cite{plant} &\cite{achasov1,achasov2} & \\
\hline
250 &   1.42   &   1.20   &   1.19   &   1.32   \\
400 &   1.48   &   1.21   &   1.18   &   1.39   \\
600 &   1.55   &   1.22   &   1.19   &   1.44   \\
 \hline
Only $\rho$, $\pi$& 9.4 & 10.4 & 10.3 &  / \\
\end{tabular}
\end{ruledtabular}
\end{table}

Next, we add the $D/S$ ratio \cite{DSratio} to the set of fitted
experimental values and repeat the calculations. The results are shown 
in Table \ref{tab:chiscomb_ds}.
\begin{table}
\caption{\label{tab:chiscomb_ds}$\chi^2$/NDF of the fits to the CMD-2
\& BaBar-LE cross section data and to the $D/S$ ratio (28 data points)}
\begin{ruledtabular}
\begin{tabular}{ccccc}
 $ \Gamma_{a_1}$ & ESK             & PB/HG           &
AK              & only $a_1$ \\
   (MeV)        & \cite{eidelman} & \cite{plant}     &
\cite{achasov1,achasov2} &          \\
 \hline
250 &   3.66   &   1.95   &   1.99   &   1.96   \\
400 &   1.98   &   1.34   &   1.33   &   1.48   \\
600 &   1.65   &   1.20   &   1.18   &   1.41   \\
%Only $\rho$, $\pi$& 10.5 & 10.4 & 10.3 &  / \\
\end{tabular}
\end{ruledtabular}
\end{table}
The salient feature of those results is a clear preference of the
highest assumed value (600~MeV) of the total $a_1$ width.

\begin{figure}
\setlength \epsfxsize{8.6cm}
\epsffile{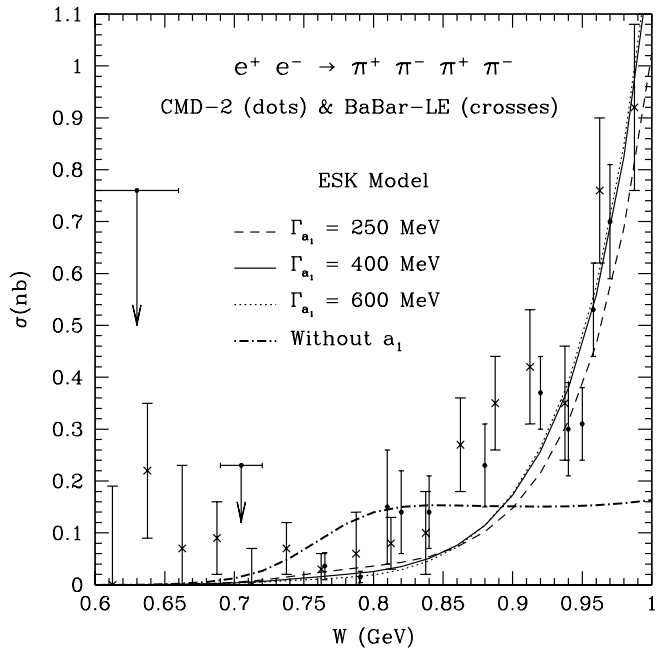}
\caption{\label{fig:esk}Excitation curves calculated in the original
(without $a_1$ meson) and expanded ESK model compared to the CMD-2 
and BaBar-LE data. The $D/S$ ratio was also used in fit.}
\end{figure}

\begin{figure}
\setlength \epsfxsize{8.6cm}
\epsffile{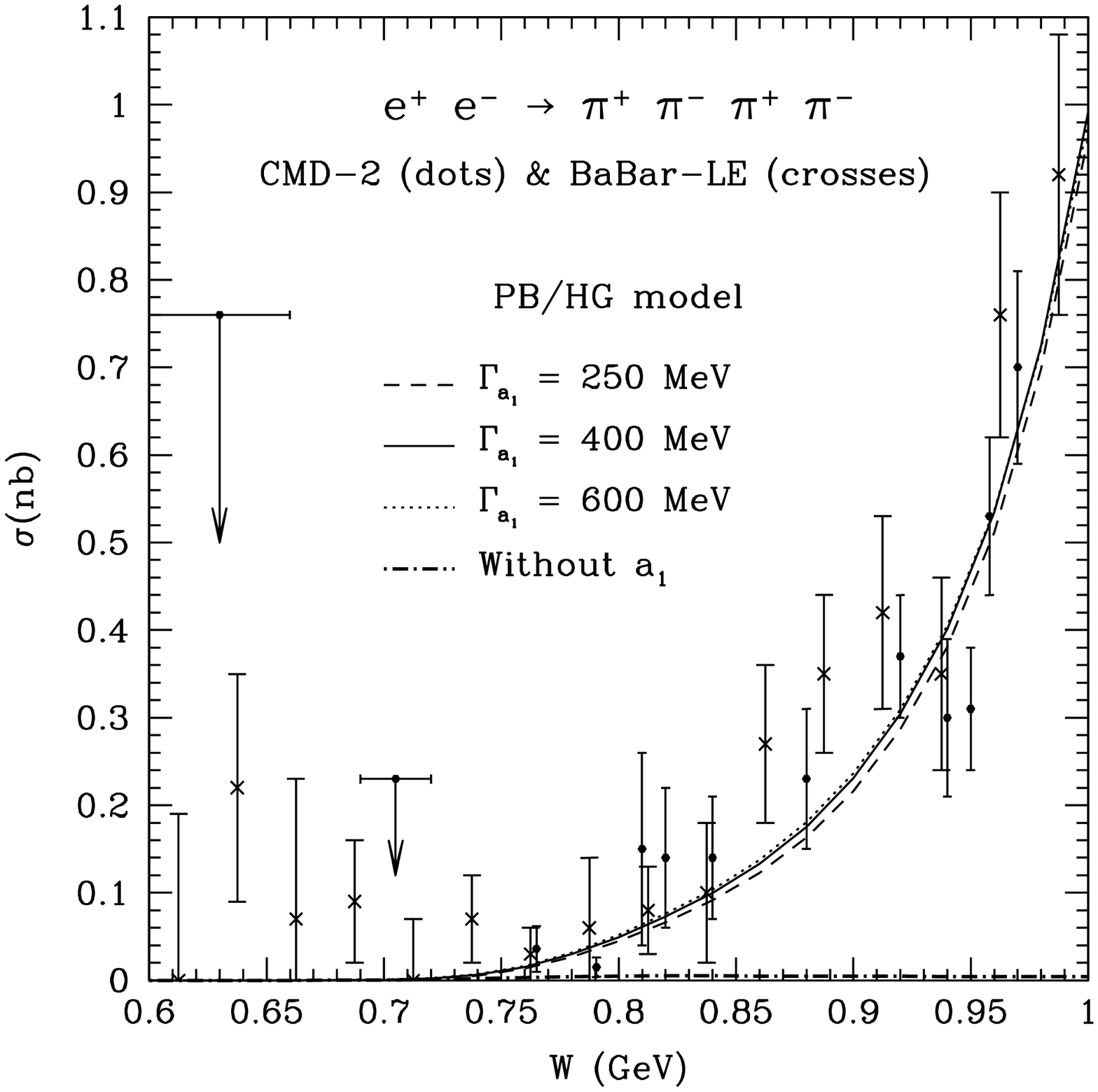}
\caption{\label{fig:pbhg}Excitation curves calculated in the original
(without $a_1$ meson; dash-dotted curve close to the abscissa) and expanded
PB/HG model compared to the CMD-2 and BaBar-LE data. The $D/S$ ratio was also
used in fit.}
\end{figure}

\begin{figure}
\setlength \epsfxsize{8.6cm}
\epsffile{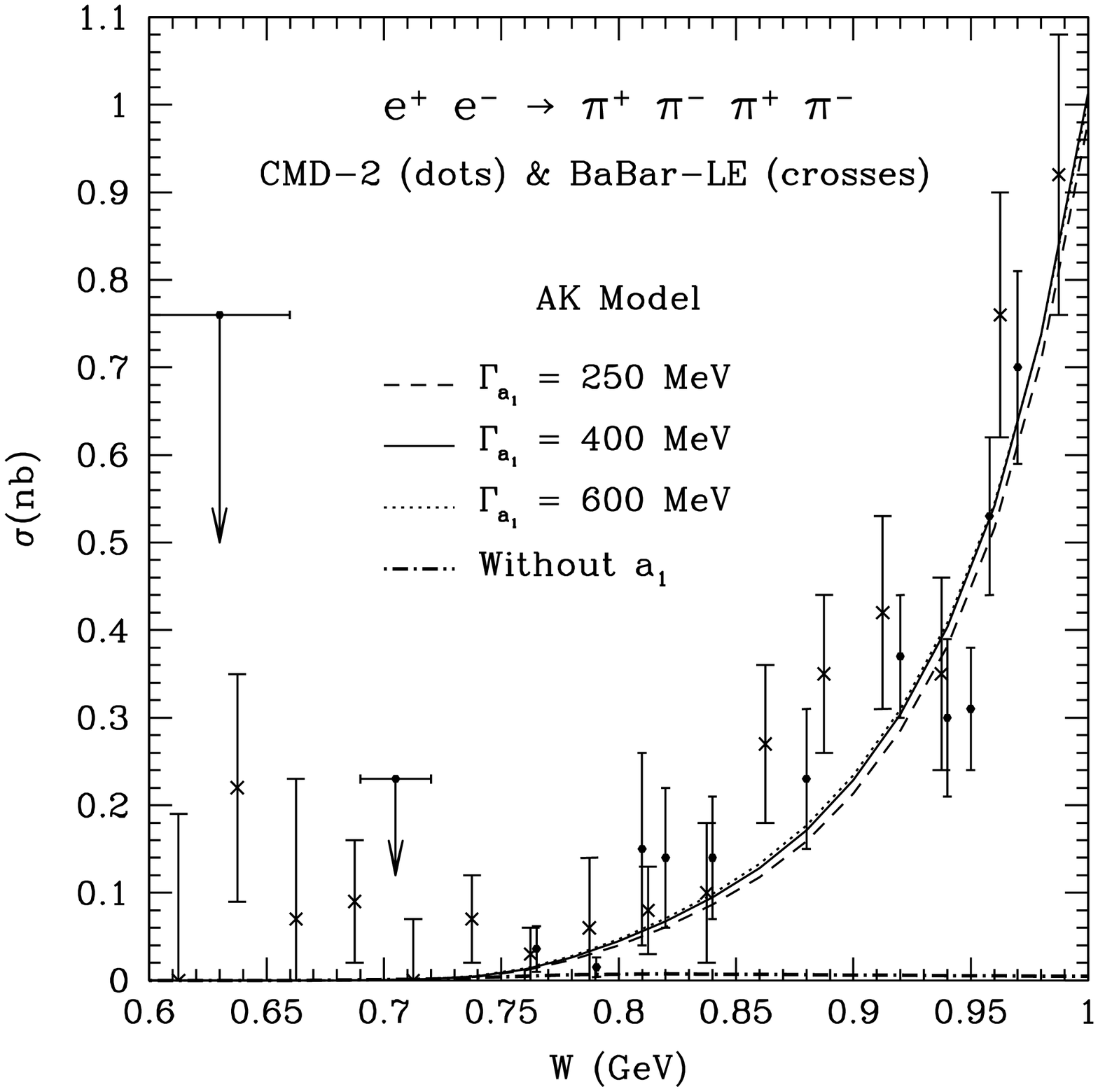}
\caption{\label{fig:ak}Excitation curves calculated in the
original (without $a_1$ meson; dash-dotted curve close to the
abscissa) and expanded AK model compared to the CMD-2 and
BaBar-LE data. The $D/S$ ratio was also used in fit.}
\end{figure}

\begin{figure}
\setlength \epsfxsize{8.6cm}
\epsffile{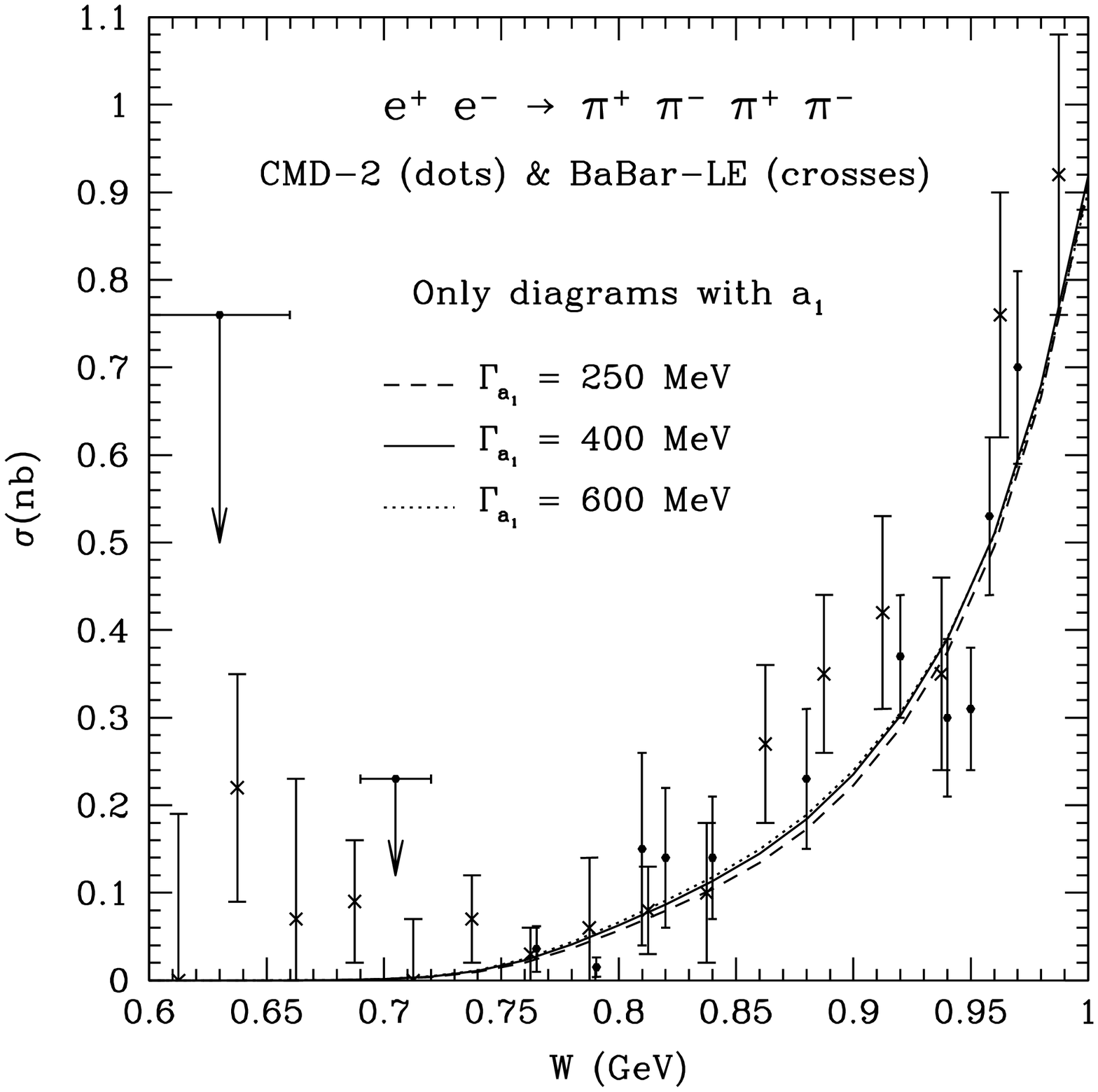}
\caption{\label{fig:a1}Excitation curves calculated from
the $a_1\pi$ diagrams only, compared to the CMD-2 and
BaBar-LE data. The $D/S$ ratio was also used in fit.}
\end{figure}

In Figs.~\ref{fig:esk}, \ref{fig:pbhg}, and \ref{fig:ak}
we show the comparison of the combined set of data with
the excitation curves calculated in all three
models supplemented with the $a_1$ diagrams Fig.~\ref{fig:diagrams}(d). 
The same comparison for the $a_1$ diagrams alone is depicted in 
Fig.~\ref{fig:a1}.

In Table~\ref{tab:sthcombds} we can see the values of
$\sin\theta$ together with their errors (defined in the usual way
\cite{minuit}) obtained from the fit to the CMD-2 \& BaBar-LE cross
section data and to the $D/S$ ratio. As we mentioned above, three
different values of the $a_1(1260)$ width are assumed. 
Table~\ref{tab:gamcombds} compares the experimental value of the 
$\rpppp$ decay width \cite{cmd2} with the results obtained from various 
models 
under the same conditions. The results for both quantities obtained with
other data sets (CMD-2 only, CMD-2 \& BaBar-LE) are very similar.

\begin{table}
\caption{\label{tab:sthcombds}Values of $\sin\theta$ from the fit to the
CMD-2 \& BaBar-LE data and to the $D/S$ ratio.}
\begin{ruledtabular}
\begin{tabular}{ccccc}
$\Gamma_{a_1}$ & ESK & PB/HG & AK & only $a_1$ \\
   (MeV) & \cite{eidelman} & \cite{plant} & \cite{achasov1,achasov2}  & \\
\hline
250 &   0.4092(33)   &   0.4278(32)   &   0.4267(32)   &   0.4312(35) \\
400 &   0.4352(24)   &   0.4624(34)   &   0.4608(32)   &   0.4679(39) \\
600 &   0.4659(27)   &   0.5046(44)   &   0.5022(41)   &   0.5132(55) \\
\end{tabular}
\end{ruledtabular}
\end{table}
\begin{table}
\caption{\label{tab:gamcombds}Decay width $\Gamma(\rpppp)$ (keV)
calculated in various models using $\sin\theta$ from the fits to the
CMD-2 \& BaBar-LE data and to the $D/S$ ratio. 
Experimental value is $(2.8\pm 1.4\pm 0.5)$~keV \cite{cmd2}.}
\begin{ruledtabular}
\begin{tabular}{ccccc}
$ \Gamma_{a_1}$ & ESK  & PB/HG & AK  & only $a_1$ \\
(MeV)    & \cite{eidelman} & \cite{plant} & \cite{achasov1,achasov2} & \\
\hline
250 &   4.28(01)   &   3.16(25)   &   2.70(23)   &   4.52(30)   \\
400 &   2.81(01)   &   3.55(28)   &   3.03(26)   &   5.08(32)   \\
600 &   1.94(02)   &   3.77(30)   &   3.22(27)   &   5.39(37)   \\
 \hline
Only $\rho$, $\pi$& 16.2   &   0.59   &   0.89  &  / \\
\end{tabular}
\end{ruledtabular}
\end{table}

\section{High-energy results ($\bm{W}$ up to 4.5~$\bm{\mathrm{GeV}}$)}
When we want to get a good description of the data on the $\die$
annihilation to four charged pions at energies above 1~GeV, we should
consider also the contribution from diagrams where higher $\rho$ 
resonances couple to the virtual photon and then convert into four
pions. We include two resonances: $\rp=\rho(1450)$ and $\rpp=\rho(1700)$.
We assume that the decay of those resonances into four pions is governed
by the same Feynman diagrams as that of $\rho(770)$, with all coupling
constants scaled by the same factor (different for $\rp$ and $\rpp$).
This assumption implies that the four-pion decay widths of $\rp$ and $\rpp$ 
have the same shape in $W$ as that of $\rho(770)$. They only differ from it
by constant factors. The simplifying assumption we have made allows us to 
use the same cross section formula \rf{convert} as in the
low energy case, with $F_{\pi}(s)$ replaced by
\be
\label{heformfactor}
F(s)=F_\rho(s)+\delta F_\rp(s)+\epsilon F_\rpp(s),
\ee
where $F_\rho(s)$ differs from \rf{piff} by including also 
$\Gamma_{\rpppp}(s)$ into the total decay width of $\rho(770)$, 
\be
F_\rp(s)=\frac{m_\rp^2}{m_\rp^2-s-im_\rp\Gamma_\rp},
\ee
and similar expression holds also for $F_\rpp(s)$. Constants $\delta$ and 
$\epsilon$ not only include the coupling constants modification factors 
mentioned above,
but also account for the fact that the couplings of $\rp$ and $\rpp$ to
photon differ from that of $\rho(770)$, which is fixed by VMD. 
Unknown complex parameters $\delta$ and $\epsilon$ will be
determined, together with the masses and widths of $\rp$ and $\rpp$ 
and other two parameters mentioned later on, by fitting the experimental 
excitation function\footnote{When we replaced \rf{piff} by \rf{heformfactor}
in the low-energy region and kept the form-factor parameters as determined 
in the high-energy fit, we got the results that differed slightly from that 
in Sec.~\ref{lowenergy}. If we varied also those parameters when fitting 
the low-energy data, they acquired unphysical values (masses of $\rp$ and 
$\rpp$ around 1~GeV). It may signify that some contribution important at 
low energies (scalar resonances) is still missing in our approach.}.

Another effect that has to be taken into account when dealing with higher
energies is connected with the structure of the strongly interacting
particles. Our decay amplitudes have been derived under the assumption that
the pions, $\rho$'s, and $a_1$'s are elementary quanta of the corresponding
quantum fields. But this assumption is justified only when their mutual
interaction is soft. When the momenta of the mesons which enters a specific
interaction vertex get higher, the contribution of that vertex to the
amplitude becomes smaller than in the case of the point-like
participants. This effect is usually described by strong form factors.
Given the present status of the strong interaction theory, we have to
turn to models. For example, in the chromoelectric flux-tube 
breaking model of Kokoski and Isgur \cite{kokoski}, the vertex describing
a two-body decay is modified by the factor $\exp\{-{p^*}^2/(12\beta^2)\}$, 
where $p^*$ is the three-momentum magnitude of the decay products in the 
parent particle rest frame and $\beta\approx 0.4$~GeV. For decay of the 
$\rho$ meson with a (non-nominal) mass $W$ into two on-mass-shell pions, 
this form factor can be written as
\be
\label{kiff}
F_{KI}(s)=\exp\left\{-\frac{s-s_0}{48\beta^2}\right\},
\ee
where $s=W^2$ and $s_0$ is the threshold value of $s$ ($4m_\pi^2$ in
the two-pion decay). The complete amplitude of the four-pion 
decay of $\rho^0$ contains many vertices, some of them with more than
three incoming/outgoing particles. Applying the Kokoski-Isgur factor
to each of them would be cumbersome and would require additional assumptions
in the case of more complicated vertices. We will therefore assign an 
``effective'' strong form factor of the form \rf{kiff} to the complete 
amplitude of the decay $\rpppp$, but with $s_0=16m_\pi^2$. When fitting 
the experimental excitation curve, $\beta$ will be considered as another 
parameter. With the masses and widths of $\rp$ and $\rpp$, complex 
parameters $\delta$ and $\epsilon$, and with the sine of the mixing angle 
$\theta$ there are ten real parameters to be determined by fitting the 
$\eepppp$ cross section data of BaBar collaboration \cite{babar}. 

In the following, we assume the $a_1$ decay width of 600~MeV,
for which the results of all models at energies below 1~GeV were best.
The same value was used in the $a_1$ propagator \rf{a1propagator}. 

The resulting optimized values of the parameters listed above
and their MINUIT \cite{minuit} errors are shown in 
Table~\ref{tabbabar} for the three different models supplemented with 
the $a_1\pi$ intermediate states and for the latter alone. Mutual 
comparison of the $\chi^2/$NDF ratios clearly
shows that the presence of the $a_1\pi$ intermediate states is crucial
for obtaining good agreement of the calculated excitation curve with
data. They provide a good fit even if taken alone. Adding the diagrams 
with $\pi$'s and $\rho$'s in the intermediate states does not change
the quality of the fit if their amplitudes are taken from the PB/HG and
AK models. On the other hand, the inclusion of the amplitudes of the 
ESK model brings some deterioration of the fit.
The values of parameter $\beta$ do not differ very much from the value 
advocated in \cite{kokoski} for the three-line vertices, what indicates 
that the effective-strong-form-factor approach we have chosen \rf{kiff} is
reasonable. The values of 
the sine of the mixing angle $\theta$ are somewhat lower than those at
low energies. The central values of the masses and widths of $\rp$ and
$\rpp$ a little different from those listed in \cite{pdg2006}, but 
differences are acceptable keeping in mind relatively large errors.

\begin{figure}
\setlength \epsfxsize{8.6cm}
\epsffile{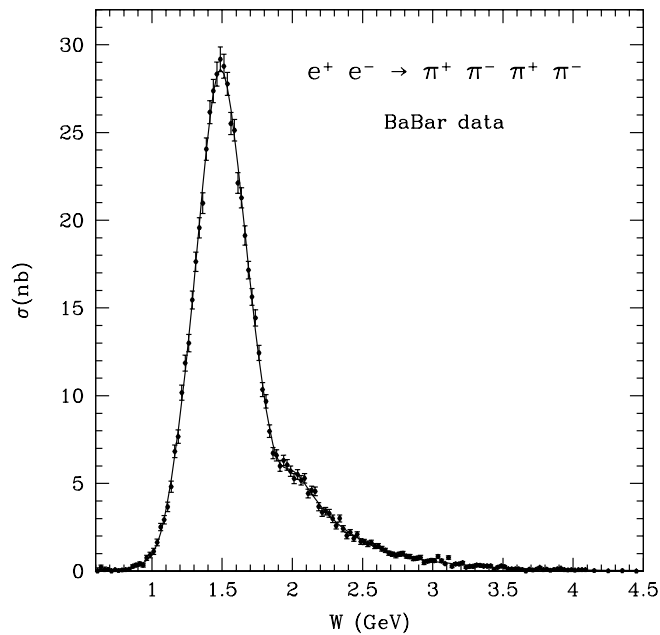}
\caption{\label{fig:babar}Theoretical excitation curve compared with 
the BaBar data \cite{babar}. The $D/S$ ratio was also used in fit. The 
result of the pure $a_1$ model is shown. The other models combined with 
$a_1\pi$ intermediate states provide almost identical curves.}
\end{figure}

The graphical comparison of the excitation curve calculated from
the model containing only the diagrams with the $a_1\pi$ intermediate 
states, shown in Fig.~\ref{fig:diagrams}(d), with experimental data 
\cite{babar} is presented in Fig.~\ref{fig:babar}.
The excitation curves of other models (ESK, PB/HG, AK) combined with the 
$a_1\pi$ contribution differ only slightly and are not shown. 

\begin{table}
\caption{\label{tabbabar}
Results of the fit to the BaBar cross section data and to the $D/S$ ratio
($145$ data points) for $\Gamma_{a_1}=600$~MeV.}
\begin{ruledtabular}
%\begin{tabular}{lll}
\begin{tabular}{ccccc}
Model & ESK   & PB/HG  & AK    & only $a_1\pi$ \\
      & \cite{eidelman} & \cite{plant} & \cite{achasov1,achasov2}  & \\
\hline
 $\chi^2/$NDF    & 1.21       & 1.12       & 1.12       & 1.12       \\
 $\sin\theta$    & 0.4474(22) & 0.4592(28) & 0.4588(27) & 0.4603(28) \\
 $\beta$ (GeV)   & 0.3505(89) & 0.3665(97) & 0.3657(97) & 0.3695(98) \\
 $m_{\rp}$ (GeV) & 1.419(12)  & 1.439(13)  & 1.438(13)  & 1.442(13)  \\
 $\Gamma_{\rp}$ (GeV)& 0.564(20)& 0.568(21)& 0.568(21)  & 0.566(21)  \\
 Re($\delta$)    & 0.1038(41) & 0.1145(51) & 0.1144(51) & 0.1149(53) \\ 
 Im($\delta$)    &-0.039(11)  &-0.019(12)  &-0.021(12)  &-0.015(13)  \\ 
 $m_{\rpp}$ (GeV) & 1.903(21) & 1.923(24)  & 1.922(24)  & 1.926(24)  \\
 $\Gamma_{\rpp}$ (GeV)& 0.247(38)& 0.284(44)& 0.283(44) & 0.290(45)  \\
 Re($\epsilon$)&-0.0016(11) &-0.0002(17) &-0.0003(17) & 0.0002(18) \\
 Im($\epsilon$)&-0.00373(94) &-0.0054(12) &-0.0054(12) &-0.0056(13)\\
\end{tabular}
\end{ruledtabular}
\end{table}

\section{Conclusions and comments}

Our low-energy results show that the inclusion of the $a_1\pi$ intermediate 
states
is of vital importance for obtaining a good agreement with the experimental 
data on the cross section of the reaction $\eepppp$ as a function of the 
incident energy (see Tables~\ref{tab:chiscmd2}, \ref{tab:chiscomb}, and 
\ref{tab:chiscomb_ds} or Figs.~\ref{fig:esk}, \ref{fig:pbhg}, and 
\ref{fig:ak}). The $\chi^2/$NDF ratio gets much smaller if a particular 
model is supplemented with the diagrams containing the $a_1$ resonance 
in the intermediate states. Viewing from another perspective, the
pure $a_1$ model provides relatively good agreement with the cross
section data, much better than each of the three models without the $a_1\pi$
intermediate states. See Tables mentioned above and Fig.~\ref{fig:a1}.
Adding the diagrams from the original models to the pure $a_1$ model improves
the fit in the case of the PB/HG and AK models, but worsens it in the
case of the ESK model. Unfortunately, the $\chi^2/$NDF ratio remains
greater than one everywhere. It may be the consequence of our ignoring some 
important contributions, perhaps those with a scalar resonance considered in 
\cite{decker,czyz}. The original models ESK, PB/HG, and AK do not contain 
any scalar resonances. We also have not included them as our main concern 
was the role of the $a_1$ resonance. It is also possible that the $\arp$
Lagrangian should contain more terms than considered in \rf{genlag}.

Using the $D/S$ ratio as an additional data point is important. It can 
discriminate among various models. In our case it increases the separation 
of the ESK model from the others. It also strongly prefers larger
values of the assumed $a_1$ width. In calculations without the $D/S$ ratio
we were able to find a value of $\sin\theta$ for each $\Gamma_{a_1}$ that 
led to an acceptable fit to the $e^+e^-\rightarrow 2\pi^+ 2\pi^-$
cross section. However, the lowest value of $\Gamma_{a_1}$ is excluded
if we add the $D/S$ ratio to the fitted data (see 
Table~\ref{tab:chiscomb_ds}).

As to the partial decay width of $\rpppp$, the conclusion is not so
categorical. Two models (PB/HG and AK) in their original forms provided 
results that were a little smaller, but did not contradict strongly the 
experimental value with its large errors. Only the original ESK model 
gave too large figure, which was in a clear disagreement with the 
experimental value. The inclusion of the $a_1\pi$ intermediate states 
brought all values into the interval given by the experimental value 
and its errors summed linearly, see Table~\ref{tab:gamcombds}. It must
be said that the pure $a_1$ model gives the decay widths that are
close to the one-sigma upper limit or even beyond it. 

We originally hoped that our study would tell us the form of the $\arp$
Lagrangian. But with respect to the $\arp$ Lagrangian, no clear picture 
can be inferred from the low-energy results yet. The optimal values of the 
sine of the mixing angle, see Table~\ref{tab:sthcombds}, are from a 
broad interval and depend not only on the choice of the original model
to which the $a_1$ diagrams are added, but also on the assumed value of 
the $a_1$ width. The optimized values of $\sin\theta$ squeeze into interval
$(0.40,0.51)$. 

The quality of the fit over the whole energy range of the BaBar experiment 
\cite{babar}, as measured by the $\chi^2/$NDF ratio, seems to be better
than that at low energies. But a more careful investigation in terms of
the confidence level, which takes into account $\chi^2$ and NDF separately,
shows equal quality of those two fits. The values of $\sin\theta$
are compatible with those found at low energies,
$\sin\theta\in(0.41,0.47)$ (lower boundary is obtained from the fits
where $\Gamma_{a_1}=250$ MeV was used). They occupy a narrower
interval, what can be explained by lesser importance of sub-dominant
diagrams with only $\rho$ and $\pi$ mesons in the intermediate states.

It is interesting to compare our estimates of the mixing parameter 
$\sin\theta$, defined in Eq.~\rf{genlag}, with its values that have been 
used so far, see Table~\ref{tab:sinus}. The values in the first and fifth 
row simply
reflect the fact that Lagrangians in Refs.~\cite{xiong,haglin,janssen}
contained just one term. The remaining rows refer to various sets
of four fundamental parameters ($m_0$, $g$, $\sigma$, and $\xi$) of the 
model in which the vector and axial-vector mesons were included as massive 
Yang-Mills fields of the SU(2)$\times$SU(2) chiral symmetry \cite{song}. 
The model built on previous works \cite{gomm84,holstein,meissner}. 
Our mixing parameter is related to the parameters $\eta_1$ and $\eta_2$ 
of that model, which can be expressed in terms of the four fundamental 
parameters using Eqs. (2.9) and (2.10) in \cite{song}. The formula is very 
simple
\[
\sin\theta=\frac{\eta_2}{\sqrt{\eta_1^2+\eta_2^2}}.
\] 
In \cite{song}, the fundamental parameters of the massive Yang-Mills model
were determined using the experimental values of the masses and width of
the $\rho$ and $a_1$ mesons. This procedure is not unique, there are
two solutions. The corresponding mixing parameter is shown in the second
and fourth row of Table~\ref{tab:sinus}. Row 3 corresponds to an 
\textit{ad hoc} choice of fundamental parameters made in \cite{turbide1}.
Last two rows show the range of our results obtained from various models
(ESK, PB/HG, AK) and various assumed values of $\Gamma_{a_1}$ (250, 400, 
and 600 GeV). Unfortunately, there is no overlap with the rows above.
The issue definitely requires more attention. The phenomenological models
may be improved by including other intermediate states, perhaps those
with scalar resonances. The parameters of the massive Yang-Mills model
may be tuned by using richer experimental input ($\Gamma(a_1\ra\pi\gamma)$
and $D/S$ ratio as in \cite{korudaz,gaogale}) and more realistic 
formulas for relating theoretical parameters to experimental quantities,
\eg, calculating the $a_1$ width as $a_1\ra 3\pi$ instead of
$a_1\ra\rho\pi$.
 
\begin{table}
\caption{\label{tab:sinus}
Survey of the mixing parameter $\sin\theta$ for various versions of
two-component $\arp$ Lagrangian \rf{genlag} that appeared in
literature.}
\begin{ruledtabular}
\begin{tabular}{ccc}
No. & $\sin\theta$ & Reference \\
\hline
1 & 0       & \cite{xiong,haglin} \\
2 & 0.2169  & \cite{song,gaogale,sokoga} \\
3 & 0.5582  & \cite{turbide1} \\
4 & 0.6308  & \cite{song,gaogale,turbide2} \\
5 & 1       & \cite{janssen} \\
\hline
  & 0.40--0.51 & our low-energy fits \\
  & 0.41--0.47 & our all-energy fits \\
\end{tabular}
\end{ruledtabular}
\end{table}

Our failure to obtain a more precise value of the mixing angle of the
$\arp$ Lagrangian suggests that it is necessary to make a simultaneous
fit to data about several physical processes. The natural candidates are
the $\die$ annihilation into various four-pion final states, the decay
of the $\tau$ lepton into neutrino and three or four pions, and the
exclusive hadronic reactions of the type investigated, \eg, in
\cite{janssen}.

\begin{acknowledgments}%
One of us (P.~L.) is indebted to David Kraus for useful discussions. We
thank Prof.~T.~Barnes for useful correspondence. This work was supported 
by the Czech Ministry of Education, Youth and Sports under contracts
MSM6840770029, MSM4781305903, and LC07050.
\end{acknowledgments}

\end{document}